\documentclass[twocolumn,showpacs,prl]{revtex4}
\usepackage{graphicx}
\usepackage{amsmath}

\begin{document}

\title{
Pfaffian pairing wave functions in electronic structure quantum Monte Carlo}

\author{M. Bajdich, L. Mitas,  G. Drobn\'y and L. K. Wagner}
\affiliation{
Center for High Performance Simulation and Department of Physics
North Carolina State University, Raleigh, NC 27695}

\author{K. E. Schmidt}
\affiliation{Department of Physics,
Arizona State University, Tempe, AZ 85287}

\date{\today}

\begin{abstract}
We investigate the accuracy of trial wave function
for quantum Monte Carlo based on pfaffian functional form with 
singlet and triplet pairing. 
Using a set of first row atoms and molecules we find that these
wave functions provide very consistent and systematic
behavior in recovering the correlation energies on the 
level of 95\%. In order to get beyond this limit we explore
the possibilities of multi-pfaffian pairing wave functions.
We show that a small number of pfaffians recovers another 
large fraction of the missing correlation energy comparable to the 
larger-scale configuration interaction wave functions. 
We also find that pfaffians lead to substantial improvements 
in fermion nodes when compared to Hartree-Fock wave functions.
\end{abstract}

\maketitle

Wave functions of
interacting quantum systems such as electrons 
in matter are notoriously difficult 
to calculate despite decades of effort. 
One of the most productive
many-body methods for electronic structure problems
is quantum Monte Carlo
 (QMC) which employs stochastic approaches
 both for solving the stationary Schr\"odinger equation
 and for evaluation of expectation values 
\cite{qmchistory,hammond,qmcrev}.
The key advantage of
QMC is its capability to use explicitly 
correlated wave functions which allow
to study many-body effects beyond the reach
of mean-field approaches. The most important limit on QMC
accuracy is the fixed-node approximation which is 
used to circumvent the fermion sign
problem \cite{jbanderson75,reynolds82}.
Even with this approximation, the fixed-node QMC
has been very successful 
for a host of real systems such as molecules, clusters 
and solids with up to 
hundreds of valence electrons and has provided an agreement 
with experiments (1-3\%) for cohesive energies, band gaps,
and other energy differences \cite{qmcrev}.
However, reaching beyond the fixed-node limit has proven to be 
 challenging since fermion nodes (subset of position space
where the wave function vanishes)
are complicated high-dimensional manifolds which are affected
by correlation as well.
Accuracy of QMC trial wave functions is therefore crucial for both
fundamental and also computational reasons. 
 
The state-of-the-art QMC calculations          
employ accurate Slater-Jastrow wave functions  
that can be written as  $ \Psi_{T}=\Psi_{A} \exp[U_{corr}]$
where $\Psi_A$ is the antisymmetric part while $U_{corr}$
describes the electron-electron and higher-order correlations.
The antisymmetric part is either 
a single Hartree-Fock (HF) determinant of one-particle orbitals
or a multi-reference sum of excited determinants such as 
a limited Configuration Interaction (CI) expansion \cite{szabo}.
A natural generalization of one-particle orbital is a two-particle
or pair orbital, sometimes called a geminal. In particular, 
the Bardeen-Cooper-Schrieffer (BCS) wave function, which 
is an antisymmetrized
product of singlet pairs, has been recently used 
to calculate several atoms and molecules as well as  
superfluid Fermi gases \cite{carlsonbcs, sorellabcs1, sorellabcs2}.
The results show significant improvements 
over the single-determinant HF wave functions. 
Nevertheless, for partially
spin-polarized systems the energy gains are
less pronounced due to the lack of pairing correlations in
the spin-polarized subspace \cite{sorellabcs1}. The spin-polarized 
triplet pairing wave functions based on pfaffians
have been tried a few times on model
systems \cite{bouchaud,Bhattacharjee, kevin}. 

In this letter, we propose to describe systems
of electrons by pfaffian wave functions 
with variational freedom
beyond HF and also BCS wavefunctions.
The pfaffian allows us to incorporate pair orbitals for 
both singlet and triplet pairing channels together with 
unpaired one-particle orbitals 
into a single, compact wave function.
These pfaffian wave functions are tested on atomic and molecular
systems in variational 
and fixed-node diffusion Monte
Carlo methods. The results show significant gains in correlation
energy both for spin-polarized and 
unpolarized cases. Furthermore, we explore the multi-pfaffian wave functions and we find that
they recover a large fraction of the missing correlation energy 
while being much more compact than expansions in determinants.

Let us consider $1,2, ...,N$ spin-up and $N+1, ...,2N$ 
spin-down electrons in a singlet
state with electron spatial coordinates 
denoted simply as $R=(1,2,...,2N)$.
An antisymmetrized product of singlet pair orbitals
 $\phi(i,j)=\phi(j,i)$
is the BCS wave function 
\begin{equation}
\Psi_{BCS}={\cal A} [\phi (i,j)] = {\rm det} [\phi (i,j)] 
={\rm det}[{\boldsymbol \Phi}]
\end{equation}
which is simply a determinant of $N\times N$ matrix.
The BCS wave function is efficient for describing
systems with single-band correlations such
as Cooper pairs in conventional BCS
superconductors where pairs form from
one-particle states close to the Fermi level.
For partially spin-polarized states one can augment the
matrix by columns/rows of one-particle orbitals.
However, the spin-polarized subspace 
is then uncorrelated and for a fully spin-polarized system
one ends up with the usual Hartree-Fock
wave function.
In order to correlate spin-polarized electrons  it is
necessary to generalize the wave function form and introduce 
effects of triplet pairing. 

For a system of $2N$ fully spin-polarized electrons
the pairing wave function is formed as
an antisymmetrized product of
triplet pair
orbitals $\chi(i,j)=-\chi(j,i)$ and is given
by \cite{bouchaud,Bhattacharjee,kevin}
\begin{equation}
{\cal A} [\chi(1,2)\chi(3,4)...] =               
{\rm pf} [\chi(i,j)]={\rm pf} [{\boldsymbol \chi}]
\end{equation}
which defines a pfaffian of degree $2N$, eg,
for $N=2$ 
\begin{equation}
{\rm pf}[\chi(i,j)]=
\chi(1,2) \chi(3,4) -\chi(1,3)\chi(2,4)+
\chi(1,4)\chi(2,3).
\end{equation}
Any pfaffian of an odd degree vanishes, however,
the pfaffian wave function can be easily generalized
to an odd number of electrons by extending 
the pfaffian by a row/column of one-particle (unpaired)
orbital. For example,
for three spin-up electrons, replace the last row/column
 in the equations above as 
$\chi(i,4) \to \varphi(i)$ and $\chi(4,i) \to -\varphi(i)$.
We note that the Hartree-Fock wave function is a special case
of both BCS singlet and pfaffian triplet pairing wave functions
as discussed \cite{subsequentpaper}. 

The square of the pfaffian is related to
the determinant of a skew-symmetric matrix as
\begin{equation}
\big\{ {\rm pf} [\chi(i,j)]  \big\} ^2 ={\rm det}[\chi(i,j)].
\end{equation}
However, the QMC applications require 
also the knowledge of
the wave function sign, eg,  for enforcing the fixed-node restriction.
Therefore, we have implemented a direct evaluation of pfaffian based
on an $O(N^3)$ algorithm which is analogous to Gauss elimination
for determinants. Note that
pfaffians can be expanded in pfaffian 
minors and by exploring Cayley's results \cite{cayley} one can
 calculate the pfaffian and its updates for
electron moves in computer time similar
to calculation of determinants \cite{subsequentpaper}.

Let us now consider a partially spin-polarized system with
unpaired electrons. 
Remarkably, the pfaffian form can accommodate
both singlet and triplet pairs as well as one-particle
unpaired orbitals into a single, compact wave function.
The singlet/triplet/unpaired (STU) orbital pfaffian wave function is given
by
\begin{equation}
\Psi_{STU}=
{\rm pf}\begin{bmatrix}
{\boldsymbol \chi}^{\uparrow\uparrow} & 
{\boldsymbol \Phi}^{\uparrow\downarrow} & 
{\boldsymbol\varphi}^{\uparrow} \\
-{\boldsymbol \Phi}^{\uparrow\downarrow T} &
{\boldsymbol \chi}^{\downarrow\downarrow} &
{\boldsymbol \varphi}^{\downarrow} \\
-{\boldsymbol\varphi}^{\uparrow T} &
-{\boldsymbol\varphi}^{\downarrow T} &
0 \;\; \\
\end{bmatrix}
\end{equation}
where the bold symbols are block matrices/vectors of
 corresponding
orbitals and $T$ denotes
transposition. For a spin-restricted STU wave function the
pair and one-particle orbitals of spin-up and -down channels
would be identical.

The pfaffian wave functions were used 
in QMC calculations
by variational and fixed-node diffusion Monte Carlo
 (VMC and DMC) methods \cite{hammond,qmcrev}.
The VMC trial/variational wave function is a product
of an antisymmetric part $\Psi_A$
times a Jastrow correlation factor
\begin{equation}\label{eq:antiwf}
\Psi_{VMC} (R) = \Psi_{A}(R) \exp[U_{corr}(\{r_{ij}\},\{r_{iI}\},
\{r_{iJ}\})]
\end{equation}
where  $U_{corr}$ depends on
electron-electron, electron-ion 
and  electron-electron-ion combinations of distances \cite{qmcrev,cyrus}
with maximum of 22 variational parameters.
For the antisymmetric part we
have used $\Psi_A=\Psi_{HF}$
and $\Psi_A=\Psi_{STU}$ as well as some tests with
$\Psi_A=\Psi_{BCS}$ to compare with recent 
results \cite{sorellabcs1, sorellabcs2}.
The pair orbitals were expanded  
in products of one-particle orbital
basis \cite{sorellabcs1}  as
\begin{equation}\label{eq:paring_orbs}
\phi(i,j), \chi(i,j)  =\sum_{k,l} c_{kl} \varphi_k(i)\varphi_l(j).
\end{equation}
The coefficients are symmetric ($c_{kl}=c_{lk}$) for
the singlet  $\phi(i,j)$, and antisymmetric ($c_{kl}=-c_{lk}$) for 
the triplet $\chi(i,j)$ functions. 
The expansions include both occupied and unoccupied
(virtual) one-particle orbitals.
The one-particle atomic and molecular
orbitals used in expansions, which we
tested, were either Hartree-Fock orbitals or natural orbitals
\cite{szabo} from 
CI correlated calculations. 
Typically, we used about 10
virtual orbitals and the natural orbitals produced better and more systematic
results than the HF ones.
The pair orbital expansion coefficients were then optimized
in VMC by minimizations of energy using recently published method
\cite{cyrus}. 

\begin{figure}[!ht]
\centering
{\resizebox{3.4in}{!}{\includegraphics{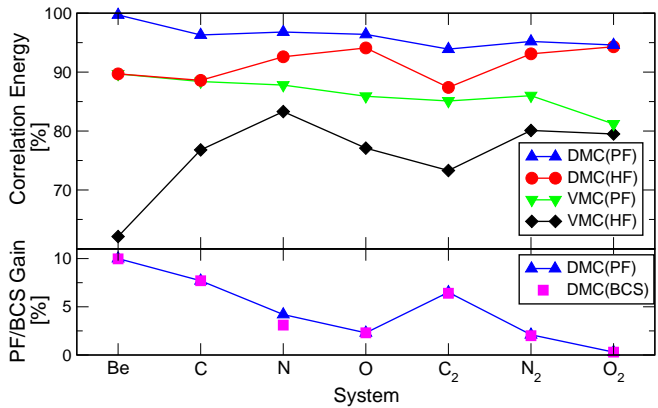}}}
\caption{
Correlation energies obtained by QMC methods
with the different trial wave functions: VMC
and fixed-node DMC with HF nodes (HF) and STU pfaffian nodes (PF).
The lower plot shows the fixed-node DMC correlation energy gains
over HF nodes for BCS and STU pfaffian
wave functions.
The statistical error bars are of the symbol sizes or smaller.
Except for the Be atom all the calculations used the same pseudopotentials
\cite{lestersbk}.
\label{fig:1}}
\end{figure}

We have applied these developments to several first row 
atoms and dimers (Fig. \ref{fig:1}). Except for the Be atom,
we used pseudopotentials to eliminate the atomic cores
\cite{lestersbk}
while the previous calculations with BCS wave functions 
were done with all electrons \cite{sorellabcs1}. Nevertheless,
our BCS wave functions produced percentages of correlation energies
rather close to the ones obtained with all electrons 
\cite{sorellabcs1}.
Perhaps the most striking result is a systematic percentage of recovered
correlation energy (94-97\%) for systems heavier than Be (see Fig. \ref{fig:1}).
The triplet contribution for
these single pfaffian STU wave functions are small, with the only
exception being the nitrogen atom, where we see a
gain of  additional  1~\% in correlation energy 
(Table \ref{energies}) when compared
to the trial wave function without triplet pairs. 
We believe that this is due to the fact that  
the nitrogen atom is quartet and therefore has 
the highest spin polarization from all studied cases. 

Our results show that
the single pfaffian form is 
capable of capturing single-band near-degeneracies
and mixing of excited states for both
spin-polarized and unpolarized systems.
\begin{table}[ht]
\caption{Total energies (a.u.) for N atom and dimer with 
amounts of correlation energy recovered, 
in VMC and DMC methods with wave functions 
as discussed in the text.  }
\vspace{0.2cm}
\begin{center}
\begin{ruledtabular}
\begin{tabular}{l c c c c }
WF & N & E$_{corr}$[\%] & N$_2$ & E$_{corr}$[\%]\\
\hline
HF                & -9.628915  & 0       & -19.44946   &  0       \\
VMC/HF            & -9.7375(1) & 83.3(1) & -19.7958(5) &  80.1(1) \\
VMC/BCS           & -9.7427(3) & 87.3(2) & -19.8179(6) &  85.2(1) \\
VMC/STU           & -9.7433(1) & 87.8(1) & -19.821(1)  &  86.0(2) \\
DMC/HF            & -9.7496(2) & 92.6(2) & -19.8521(3) &  93.1(1) \\
DMC/BCS           & -9.7536(2) & 95.7(2) & -19.8605(6) &  95.1(1) \\
DMC/STU           & -9.7551(2) & 96.8(1) & -19.8607(4) &  95.2(1) \\
Exact/est.        & -9.759215  & 100  & -19.88196   & 100 \\
\end{tabular}
\end{ruledtabular}
\end{center}
\label{energies}
\end{table}
Considering multi-determinantal expansions, such as the CI method,
the overall trade-off between accuracy and 
computational cost seems to be in favor of more compact and
physically-based pfaffian wave functions. A similar opinion was also
expressed by Sorella and coworkers \cite{sorellabcs1}.

In order to test the limits of pfaffian functional form,
 we propose a simple extension:  a multiple pfaffian (MPF)
 wave function having a form
\begin{equation}
\Psi_{MPF}={\rm pf}[\chi_1,\phi_1,\varphi_1]+{\rm pf}[\chi_2,\phi_2,\varphi_2]+\ldots
\end{equation}
so that in Eq.~\ref{eq:antiwf} we have $\Psi_A=\Psi_{MPF}$.
In actual calculations we start with all pairing 
functions such that each pfaffian is equal to the 
HF wave function, ${\rm pf}[\chi_i,\phi_i,\varphi_i]=\Psi_{HF}$. 
The pairing orbitals [see Eq.~(\ref{eq:paring_orbs})] are 
expanded in the basis of HF/natural occupied orbitals, eg, for the carbon atom
we have $2s$, $2p_x$ and $2p_y$. The choice of singlet
$\phi_1(1,2)=2s(1)2s(2) \equiv \phi_1[2s,2s]$ and triplet 
$\chi_1(1,2)=2p_x(1)2p_y(2)-2p_y(1)2p_x(2) \equiv \chi_1[2p_x,2p_y]$ 
pairing orbitals then 
gives ${\rm pf}[\chi_1,\phi_1]=\Psi_{HF}[2s^{\uparrow\downarrow},2p_x^{\uparrow},2p_y^{\uparrow}]$. 
However, one can construct the equivalent combinations of pairs as: 
$\phi_2[2s,2p_x]$, $\chi_2[2s,2p_y]$ and  $\phi_3[2s,2p_y]$, $\chi_3[2s,2p_x]$.
We can therefore include all three 
pfaffians into our $\Psi_{MPF}$ and further optimize independently all the pairing functions 
in VMC on the space of occupied and virtual orbitals. This construction 
allows us to incorporate the excitations which are not present in the single pfaffian
STU wavefunction
based only on a single $\Phi$ and a single $\chi$ pairing functions.
In the leading order the resulting MPF wave function then corresponds 
to the CI wave function with singles and doubles with the same active orbital space.
The disadvantage of this approach is that we perform the 
VMC optimizations of $M^2$ pairing coefficients for each pfaffian given that $M$ is the 
total size of our orbital basis. 
However, this can be improved by a factor of $M$
if we 
rotate the one-particle orbitals to make the pairing functions diagonal.
The total number of pfaffians in the expansion 
is then subject to required symmetry of state and desired accuracy. In the most general
case it would be proportional to the
number of distinct pairs. For large systems one can restrict 
the number of pairs by including only the ones 
which have significant contributions, eg, pairs formed from
states close to the Fermi level, or considering pairs only within
 a given band or a sub-band.

\begin{table}[ht]
\caption{Percentages of correlation energies recovered 
for C, N and O atoms by VMC and DMC methods 
with wave functions as discussed in the text.
Corresponding number of pfaffians/determinants $n$ for each
wave function is also shown. The estimated exact correlation energies for C,N,O are 
 0.1031, 0.1303, 0.1937 a.u. \cite{dolgcpl}.}
\vspace{0.2cm}
\begin{center}
\begin{ruledtabular}
\begin{tabular}{l c c c c c c}
WF       & $n$ & C & $n$ & N & $n$ & O\\
\hline
VMC(MPF) & 3   & 92.3(1) & 5   & 90.6(1) &  11  & 93.6(2) \\ 
VMC(CI)  & 98  & 89.7(4) & 85  & 91.9(2) & 136  & 89.7(4) \\ 
DMC(MPF) & 3   & 98.9(2) & 5   & 98.4(1) &  11  & 97.5(1) \\ 
DMC(CI)  & 98  & 99.3(3) & 85  & 98.9(2) & 136  & 98.4(2) \\
\end{tabular}
\end{ruledtabular}
\end{center}
\label{energies2}
\end{table}
The results (Table \ref{energies2}) show
that our MPF wave functions are able to
recover close to  99\% of correlation energy.
Furthermore, comparison with the CI results shows that
it is possible to obtain similar quality of wave functions with corresponding
improvements of the fermion nodes
at much smaller calculational cost. This is another indication 
that the inclusion of singlet and triplet pairing
enables us to treat the correlation in both spin-polarized and -unpolarized
channels in a consistent manner.

\begin{figure}[ht]
\centering
{\resizebox{2.3in}{!}{\includegraphics{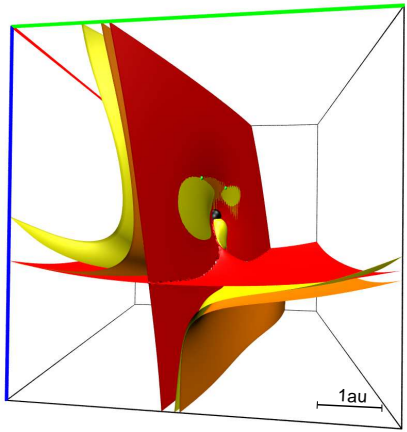}}}
\caption{
A 3D cut through the fermion node hypersurface of oxygen atom 
obtained by scanning the wave function with a pair of spin-up and -down of electrons,
both sitting at the scanning point.
The remaining electrons are fixed 
at a given VMC snaphsot positions (small green spheres). 
Nucleus is depicted in the center of the cube by the black 
sphere. The three colors show nodes of: Hartree-Fock (red/dark gray);
STU pfaffian nodes 
(orange/medium gray); and the nodes of the CI
 wave function (yellow/light gray).
The CI node is very close to the exact one (see text).
The HF node clearly divides the space into four
nodal cells while pfaffian and CI wave functions
partitioning leads to the minimal number of two nodal cells.
\label{fig:3}}
\end{figure}

The quality of fermion nodes is crucial for accurate energies
in the fixed-node DMC. 
The fermion node manifold is defined by an implicit equation
$\Psi(R)=0$. For $N$ electrons the manifold has 
 $(3N-1)$ dimensions and 
divides the configuration space into
compact nodal cells. The HF 
 spin-polarized ground states typically show two nodal 
cells, while HF unpolarized or partially polarized states have
$2\times2=4$ nodal cells since the wave function
is a product of spin-up and -down
determinants \cite{davidnode,reynoldsnode, Bressanininew}.
Changes in the HF nodal structures from inclusion of correlation
were recently investigated  by D. Bressanini {\it et al.} 
\cite{reynoldsnode,Bressanininew}.
Here we observe that the pairing correlations have an important effect
on the nodal structure as well. A changes in the nodal manifold topology
is illustrated in Fig.~\ref{fig:3} on the example of oxygen atom.
As expected, 
the HF nodes show four nodal cells
while the pfaffian "opens" these
artificial compartments and changes the topology to
the minimal number of two nodal cells, similar to the effect
of correlation observed for Be atom \cite{reynoldsnode}.
A comparison of pfaffian nodes
with very accurate CI nodes shows that both are qualitatively
similar.   Further results on the properties of nodes and their
improvement are given elsewhere 
\cite{subsequentpaper,subsequentlubospaper}.

In conclusion, we have proposed pfaffians with
singlet pair, triplet pair and unpaired
orbitals as variationally rich and compact wave functions
which offer significant and systematic improvements
over commonly used Slater determinant-based
wave functions. We have demonstrated that 
these pfaffian pairing wave functions 
are able to capture a large fraction of
missing correlation energy with consistent treatment of
correlation for both spin-polarized and unpolarized pairs.
We have explored also multi-pfaffian wave functions
which enable us to obtain more correlation while keeping
the wave functions compact.
Our pfaffian wave function results represent only the lower 
bound on the recovered correlation energies since we were
focused on systematic trends using new functional forms.
For example, a simultaneous reoptimization of one-particle
orbitals used
 in representation of pairs could
improve the results further. 
Finally, the pfaffian wave functions exhibit 
qualitative improvements
in fermion nodes and eliminate a significant 
portion of the Hartree-Fock node errors.  

We gratefully acknowledge the support by ONR-N00014-01-1-0408,
and NSF DMR-0121361, DMR-0121361, and EAR-0530110 grants and
the computer time allocations at NCSA, PSC and SDSC facilities.

\end{document}